\documentclass{PoS}

\usepackage{graphicx}
\usepackage{amsmath}
\usepackage{caption}
\usepackage{subcaption}
\usepackage{algorithmicx,algpseudocode}
\usepackage{algorithm}
\usepackage{placeins}
\usepackage[export]{adjustbox}

\newcommand{\tsep}{\mathop{t_{\rm sep}}\nolimits}
\newcommand{\tsepi}{\mathop{t_{\rm sep} \to \infty}\nolimits}

\newcommand{\GeV}{\mathop{\rm GeV}\nolimits}

\newcommand{\lsim}{\raisebox{-0.7ex}{$\stackrel{\textstyle <}{\sim}$ }}

\title{Nucleon Charges from 2+1+1-flavor HISQ and 2+1-flavor clover lattices}

\ShortTitle{Nucleon Isovector Charges}

\author{\speaker{Rajan Gupta}\footnote{LA-UR-16-29008} \\
        Los Alamos National Laboratory, Los Alamos, NM, 87545, U.S.A.\\
        E-mail: \email{rajan@lanl.gov}} 
        
\author{PNDME and NME Collaborations\thanks{Calculations on the 2+1+1-flavor HISQ lattices are
    being done in collaboration with T. Bhattacharya, V. Cirigliano,
    Y. C. Jang, H-W. Lin and B. Yoon. Calculations on the 2+1-clover
    ensembles are being done in collaboration with T. Bhattacharya, V. Cirigliano,
    J. Green, B\'alint Jo\'o, Y. C. Jang, H-W. Lin, K. Orginos, D. Richards, 
    S. Syritsen, F. Winter and B. Yoon.}}

\abstract{Precise estimates of the nucleon charges $g_A$, $g_S$ and
  $g_T$ are needed in many phenomenological analyses of SM and BSM
  physics.  In this talk, we present results from two sets of
  calculations using clover fermions on 9 ensembles of 2+1+1-flavor HISQ
  and 4 ensembles of 2+1-flavor clover lattices. We show that high statistics can be
  obtained cost-effectively using the truncated solver method with
  bias correction and the coherent source sequential propagator
  technique.  By performing simulations at 4--5 values of the
  source-sink separation $\tsep$, we demonstrate control over
  excited-state contamination using 2- and 3-state fits. Using the
  high-precision 2+1+1-flavor data, we perform a simultaneous fit in $a$, $M_\pi$
  and $M_\pi L$ to obtain our final results for the charges. }

\FullConference{34th annual International Symposium on Lattice Field Theory\\
		24-30 July 2016\\
		University of Southampton, UK}

\begin{document}

\section{Introduction}

In this talk, we highlight three areas in which significant progress
has been made to extract matrix elements of quark bilinear operators
within nucleon states. (i) cost-effective increase of statistics using
the truncated solver method with bias correction and the coherent
source sequential propagator technique; (ii) inclusion of up to 4-states
(3-states) in the analysis of 2-point (3-point) correlation functions;
and (iii) a simultaneous fit in $a$, $M_\pi$ and $M_\pi L$ to data at
different $a$, $M_\pi$ and $L$ to get the physical value. The results 
presented here are based on Refs.~\cite{Yoon:2016dij,Bhattacharya:2016zcn,Yoon:2016clover}. 
A summary of the lattice ensembles used
and measurements made in the clover-on-HISQ study is given in
Table~\ref{tab:HISQ} and in the clover-on-clover study in
Table~\ref{tab:clover}.  Associated results for the isovector form factors:
$G_E(q^2)$, $G_M(q^2)$,$G_A(q^2)$, and $G_S^s(Q^2)$ were presented by
Yong-Chull Jang at this conference~\cite{Jang:2016}.

\section{Increasing Statisitics Cost-effectively} 

The various systematic uncertainties in the calculation of matrix
elements (ME) of local quark bilinear operators within nucleon states
are at the 5\% level~\cite{Bhattacharya:2016zcn}. In order to isolate,
understand and address these systematics, one needs data with
statistical errors that are significantly smaller.  To reduce
statistical errors, one needs to make measurements on significant
numbers of decorrelated lattices that adequately importance sample the
phase space of the path integral. We have found the following three
techniques to be cost-effective ways of reducing the statistical
errors.

Lattices with $M_\pi L \ge 4$ can be regarded as consisting of a large
number of uncorrelated regions, i.e., measurements of nucleon
correlation functions in different sub-regions are statistically
uncorrelated. Since the generation of lattices with dynamical fermions
is expensive, one should assess the dimensions of these sub-regions.
For ME within nucleon states we find that $O(100)$ measurements per
lattice of size $M_\pi L =4$ are 
cost-effective~\cite{Yoon:2016dij}. Furthermore, choosing the source
points randomly within these sub-regions of a lattice and between
lattices reduces correlations.

Computer time can be reduced significantly by using the coherent
source sequential propagator method~\cite{Yoon:2016dij}. For example,
if $N_{\rm meas}$ measurements on a given lattice are done in a single
computer job, then the number of quark propagators that need to be
calculated using coherent sources reduces to $N_{\rm meas} + 2$ 
from $N_{\rm meas} + 2\times N_{\rm meas}$ in the standard approach. 

Lastly, in a given measurement, one has the freedom to choose the
precision with which quark propagators are calculated. The truncated
solver method~\cite{Bali:2009hu} with bias
correction~\cite{Blum:2012uh} significantly reduces the
computational time. In Ref.~\cite{Bhattacharya:2016zcn}, we show that
a stopping residue $r_{\rm LP} \equiv |{\rm residue}|_{\rm LP}/|{\rm
  source}| = 10^{-3}$ reduces cost by a factor of about 17 compared to
$r_{\rm HP} = 10^{-10}$. Possible
bias is corrected for using
\begin{equation}
 C^\text{imp}  = \frac{1}{N_\text{LP}} \sum_{i=1}^{N_\text{LP}} 
    C_\text{LP}(\mathbf{x}_i^\text{LP}) \nonumber 
  + \frac{1}{N_\text{HP}} \sum_{i=1}^{N_\text{HP}} \left[
    C_\text{HP}(\mathbf{x}_i^\text{HP})
    - C_\text{LP}(\mathbf{x}_i^\text{HP})
    \right] \,,
  \label{eq:2-3pt_AMA}
\end{equation}
where $C_\text{LP}$ and $C_\text{HP}$ are the 2- and 3-point
correlation functions calculated in low- (LP) and high-precision (HP),
respectively, and $\mathbf{x}_i^\text{LP}$ and
$\mathbf{x}_i^\text{HP}$ are the two kinds of source positions.  Bias,
given by the second term, if present, was much smaller than the
statistical errors.  Speedup is achieved because we need 1 HP and LP measurement for bias 
correction for every 32 LP used for statisitcs.  

\begin{table}
  \renewcommand{\arraystretch}{1.2} 
  \resizebox{0.98\linewidth}{!}{
    \begin{tabular}{l|ccc|cc|ccc}
      \hline
      Ensemble ID & $a$ (fm) & $M_\pi^{\rm sea}$ (MeV) & $M_\pi^{\rm val}$ (MeV) & $L^3\times T$    & $M_\pi^{\rm val} L$ & $N_\text{conf}$  & $N_{\rm meas}^{\rm HP}$  & $N_{\rm meas}^{\rm AMA}$  \\
      \hline
      a12m310  & 0.1207(11) & 305.3(4) & 310(3) & $24^3\times 64$ & 4.55 & 1013 & 8104  &   64832   \\
      a12m220S & 0.1202(12) & 218.1(4) & 225(2) & $24^3\times 64$ & 3.29 & 1000 & 24000 &           \\
      a12m220  & 0.1184(10) & 216.9(2) & 228(2) & $32^3\times 64$ & 4.38 & 958  & 7664  &           \\
      a12m220L & 0.1189(9)  & 217.0(2) & 228(2) & $40^3\times 64$ & 5.49 & 1010 & 8080  &  68680   \\
      \hline                                      
      a09m310  & 0.0888(8)  & 312.7(6) & 313(3) & $32^3\times 96$ & 4.51 & 881  & 7048  &           \\
      a09m220  & 0.0872(7)  & 220.3(2) & 226(2) & $48^3\times 96$ & 4.79 & 890  & 7120  &           \\
      a09m130  & 0.0871(6)  & 128.2(1) & 138(1) & $64^3\times 96$ & 3.90 & 883  & 7064  &  84768   \\
      \hline                                      
      a06m310  & 0.0582(4)  & 319.3(5) & 320(2) & $48^3\times 144$& 4.52 & 1000 & 8000  &  64000   \\
      a06m220  & 0.0578(4)  & 229.2(4) & 235(2) & $64^3\times 144$& 4.41 & 650  & 2600  &  41600   \\
      a06m135  & 0.0568(1)  & 135.5(2) & 136(2) & $96^3\times 192$& 3.74 & 229  & 1145  &  36640   \\
      \hline
    \end{tabular}
  }
\caption{Summary of the ensembles used in the clover-on-HISQ
  study. For the a06m130 ensemble, preliminary results on form
  factors only were presented by Y-P Jang at this conference.}
\label{tab:HISQ}
\end{table}

\begin{table}
  \renewcommand{\arraystretch}{1.2} 
  \resizebox{0.95\linewidth}{!}{
    \begin{tabular}{l|cc|cc|ccc}
      \hline
      Ensemble ID & $a$ (fm) & $M_\pi^{\rm sea}$ (MeV) & $L^3\times T$    & $M_\pi^{\rm val} L$ & $N_\text{conf}$  & $N_{\rm meas}^{\rm HP}$  & $N_{\rm meas}^{\rm AMA}$  \\
      \hline
      a127m285  & 0.127(2) & 285(3) & $32^3\times 96$  & 5.85 & 1000 & 4020 &   128480 \\
      a094m280  & 0.094(1) & 278(3) & $32^3\times 64$  & 4.11 & 1005 & 3015 &  96480 \\
      \hline
      a091m170  & 0.091(1) & 166(2) & $48^3\times 96$  & 3.7  & 629  & 2516 & 80512 \\
      a091m170L & 0.091(1) & 172(6) & $64^3\times 128$ & 5.08 & 467  & 2335 & 74720 \\
      \hline
    \end{tabular}
  }
\caption{Summary of the ensembles used in the clover-on-HISQ study. Here $M_\pi^{\rm sea} = M_\pi^{\rm val}$ } 
\label{tab:clover}
\end{table}

\section{Reducing Systematic Uncertainties} 

The above techniques allowed us to make, on each ensemble, $O(10^5)$
measurements on $O(1000)$ lattices.  In these measurements of 2- and
3-point correlation functions, two systematics need to be addressed in
order to extract the charges, form factors, and other ME:
removing excited-state contamination (ESC) and precise determination of the renormalization factor
connecting the lattice operator to some phenomenological scheme such
as the $\overline{\text{MS}}$ scheme at $\mu = 2\GeV$.

ESC arises because typical interpolating operators used to create and
annihilate meson and baryon states on the lattice couple to the ground
state, all its excitations and multiparticle states with the same
quantum numbers.  Thus, contributions of all higher states have to be
removed to obtain the desired ME within the ground state. The behavior
of the 2- and 3-point correlation functions, given by the spectral
decomposition, has contributions from a tower of intermediate states
in terms of unknown amplitudes ${\cal A}_i$, energies $M_i$ and ME
$\langle f | \mathcal{O}_\Gamma | i \rangle$ that are extracted from
fits.  Formally, the ground state ME $\langle 0 | \mathcal{O}_\Gamma | 0 \rangle$ 
can be obtained by calculating
the 3-point functions with very large source-sink separation $\tsep$. For
baryons, the signal in 2- and 3-point function degrades exponentially, and
for values of $\tsep$ accessible with $O(10^5)$ measurements, the ESC
is found to be significant. Thus, methods to miminize ESC in
calculations with $1 \lsim \tsep \lsim 1.5$~fm are needed. \looseness=-1

\begin{figure*}[tb]
\centering
      \includegraphics[width=0.95\linewidth,trim={0      0.01cm 0 0},clip]{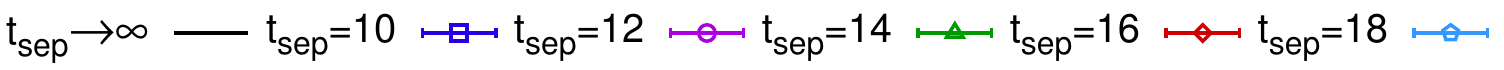}  
\\
\vspace{-0.2cm}
    \includegraphics[width=0.49\linewidth]{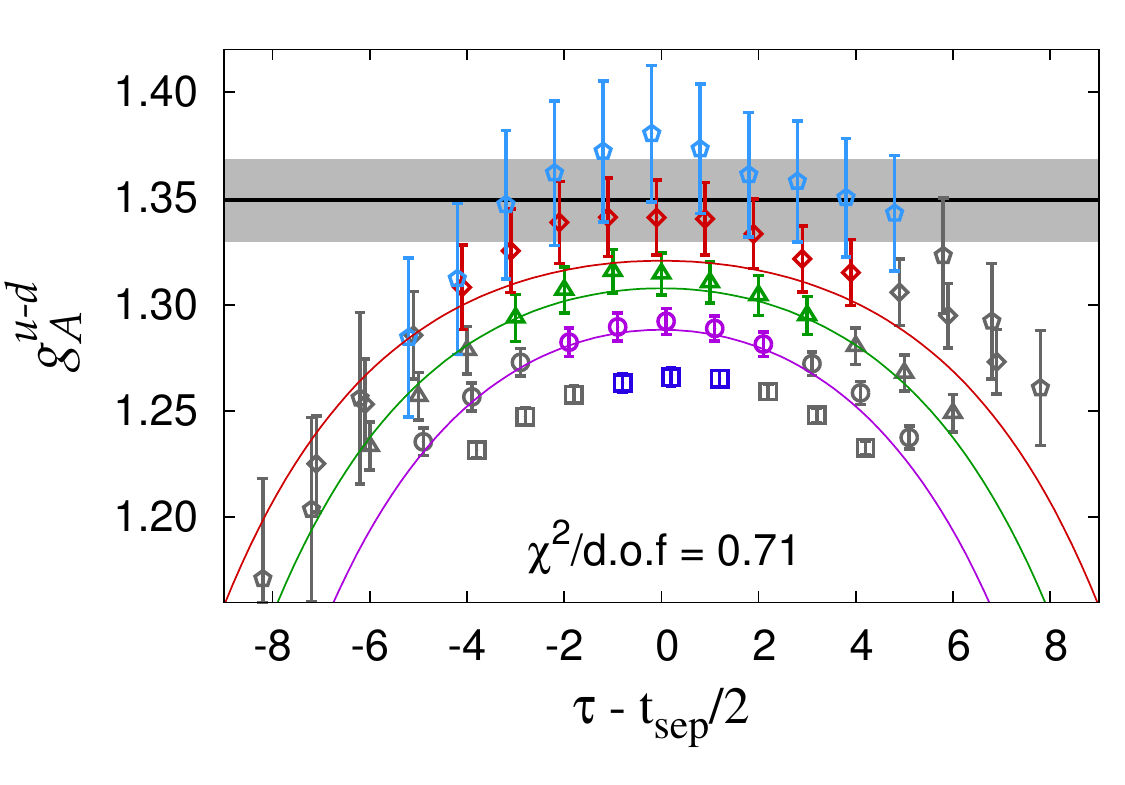}
    \includegraphics[width=0.49\linewidth]{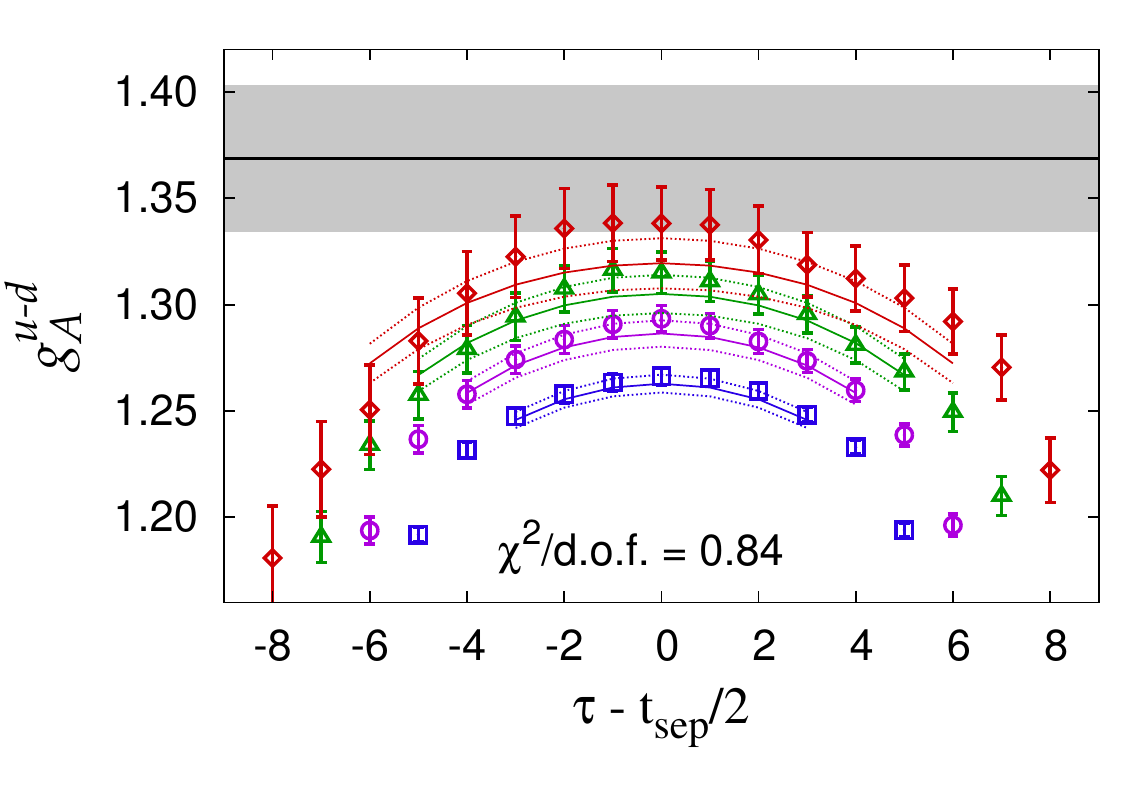}
\caption{Unrenormalized $g_A^{u-d}$ on the $a081m315$ 2+1-flavor
  clover ensemble with $\sigma = 5$ smearing.  The left (right)
  panel shows the data, 2-state (3-state) fit, and $\tsepi$ estimate (grey band).
  \label{fig:comparegAfits}}
\end{figure*}

We demonstrate control over ESC by (i) constructing the interpolating
operators using tuned smeared sources in the generation of the quark
propagators; (ii) performing the calculation at multiple values of
$\tsep$; (iii) inserting the operator at all intermediate timeslices
$\tau$ between the source and sink;  (iv) analyzing the 2- and 3-point correlators by including 
increasing number of intermediate states. For example, the 2-state 
truncation of the zero-momentum correlation functions is 
\begin{align}
C^\text{2pt}(t_f,t_i) = &{|{\cal A}_0|}^2 e^{-M_0 (t_f-t_i)} + {|{\cal A}_1|}^2 e^{-M_1 (t_f-t_i)}\,, 
\label{eq:2pt}  \\
C^\text{3pt}_{\Gamma}(t_f,\tau,t_i) =  &|{\cal A}_0|^2 \langle 0 | \mathcal{O}_\Gamma | 0 \rangle  e^{-M_0 (t_f-t_i)} +
   |{\cal A}_1|^2 \langle 1 | \mathcal{O}_\Gamma | 1 \rangle  e^{-M_1 (t_f-t_i)} +{}\nonumber\\
  & {\cal A}_0{\cal A}_1^* \langle 0 | \mathcal{O}_\Gamma | 1 \rangle  e^{-M_0 (\tau-t_i)} e^{-M_1 (t_f-\tau)} + 
   {\cal A}_0^*{\cal A}_1 \langle 1 | \mathcal{O}_\Gamma | 0 \rangle  e^{-M_1 (\tau-t_i)} e^{-M_0 (t_f-\tau)} ,
\label{eq:3pt}
\end{align}
where $\tau$ is the operator insertion time and $t_f -t_i =
t_\text{sep}$ in the 3-point function calculation.  The states
$|0\rangle$ and $|1\rangle$ represent the ground and ``first excited''
nucleon states, respectively. In 2-state analysis, the four
parameters, $M_0$, $M_1$, ${\cal A}_0$ and ${\cal A}_1$ are estimated
first from fits to the 2-point data and then used as input in fits to
3-point functions to obtain the three ME $\langle 0 |
\mathcal{O}_\Gamma | 0 \rangle$, $\langle 0 | \mathcal{O}_\Gamma | 1
\rangle $ and $ \langle 1 | \mathcal{O}_\Gamma | 1 \rangle$.  The
estimate of the charge $g_\Gamma = \langle 0 | \mathcal{O}_\Gamma | 0 \rangle$ 
improves with number of $\tsep$, the precision of the data, and the number of states included in the fits.  We
find that with $O(10^5)$ measurements, fits with 4 states (3 states)
to the 2-point (3 point) functions with full covariance matrix can be
made. Stable and consistent estimates of the charges in the $\tsepi$
limit are obtained using data with 4--5 values of $\tsep$ in the range
1--1.5~fm.  A comparison of the 2- and 3-state fits,  and the 
consistency of the $\tsepi$ value obtained for the isovector axial charge
$g_A^{u-d}$ is illustrated in Fig.~\ref{fig:comparegAfits}. \looseness=-1

Results for the various ME are then renormalized by multiplicative
factors $Z_\Gamma$ calculated using the RI-sMOM scheme as discussed in
Ref.~\cite{Bhattacharya:2016zcn}. Errors in the ME and $Z_\Gamma$ are
combined in quadratures. This gives us a set of renomalized
lattice estimates as functions of $a$, $M_\pi$ and $M_\pi L$.

\section{Simultaneous fit in $a$, $M_\pi$ and $M_\pi L$} 

\begin{figure*}[tb]
    \includegraphics[width=0.98\linewidth]{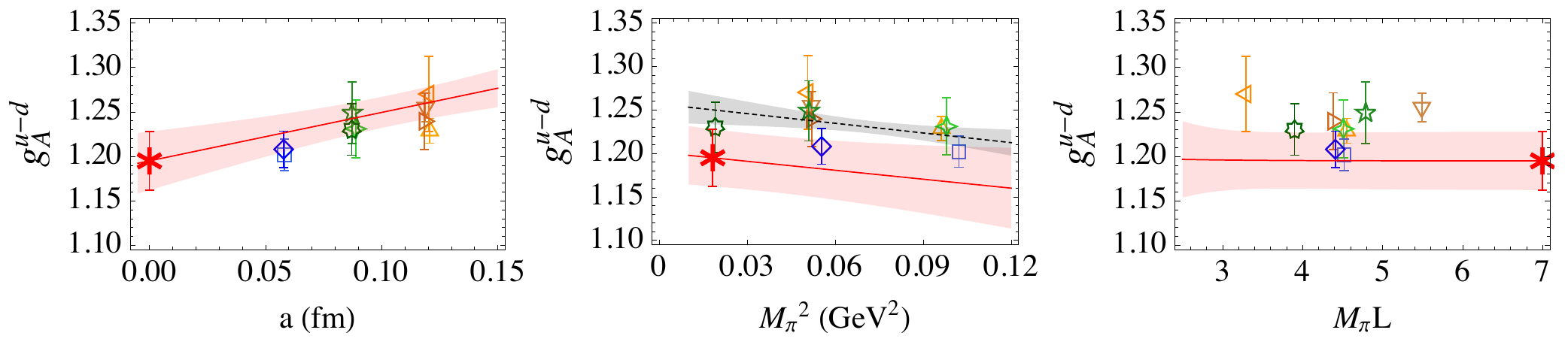}
    \includegraphics[width=0.98\linewidth]{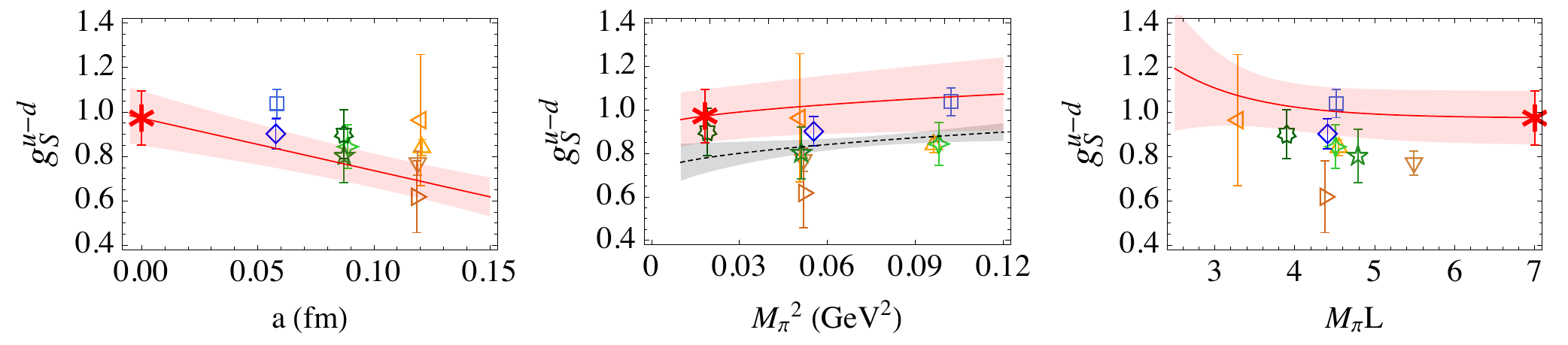}
    \includegraphics[width=0.98\linewidth]{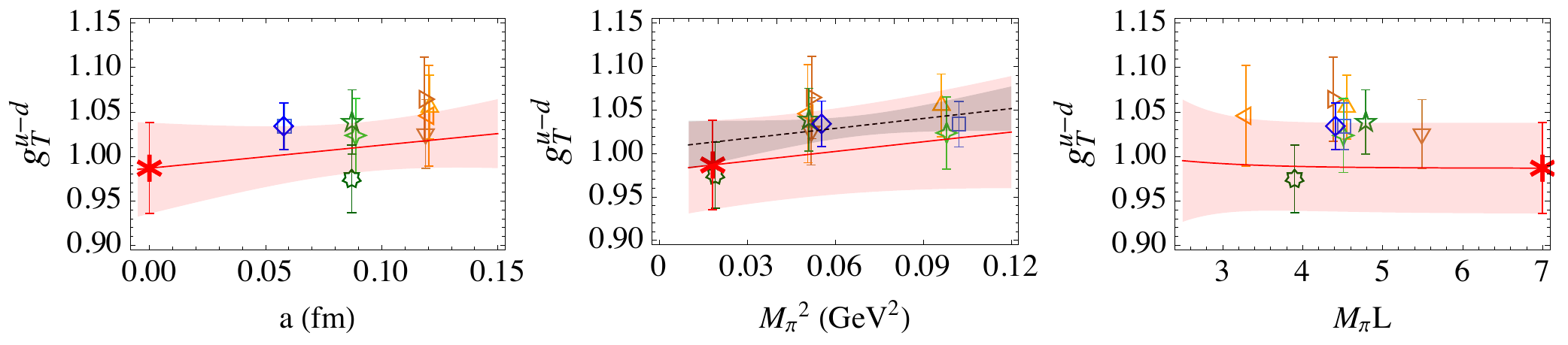}
\caption{The 9-point fit using Eqs.~\protect\eqref{eq:CextrapgAT}
  and~\protect\eqref{eq:CextrapgS} to the data for the renormalized
  isovector charges, $g_A^{u-d}$, $g_S^{u-d}$ and $g_T^{u-d}$, in the
  $\overline{{\rm MS}}$ scheme at $2\GeV$.  The result of the
  simultaneous extrapolation to the physical point defined by
  $a\rightarrow 0$, $M_\pi \rightarrow M_{\pi^0}^{{\rm phys}}=135$~MeV
  and $L \rightarrow \infty$ are marked by a red star.  The error
  bands in each panel show the simultaneous fit as a function of a
  given variable holding the other two at their physical value.  The
  data are shown projected on to each of the three planes.  The
  overlay in the middle figures with the dashed line within the grey
  band, is the fit to the data versus $M_\pi^2$ neglecting dependence
  on the other two variables.  
  \label{fig:conUmD_extrap9}}
\end{figure*}

With the renormalized estimates, calculated as functions of $a$, $M_\pi$ and $M_\pi
L$, in hand, results in the limit $a \to 0$, $M_\pi=135$~MeV and $M_\pi L \to
\infty$, are obtained using a simultaneous fit in the three
variables. With the current 9 clover-on-HISQ data points, fits are sensitive to only the lowest
order correction terms ~\cite{Bhattacharya:2016zcn}:
\begin{align}
  g_{A,T} (a,M_\pi,L) &= c_1 + c_2a + c_3 M_\pi^2 + 
 c_4 M_\pi^2 {e^{-M_\pi L}} \,,
\label{eq:CextrapgAT} \\
  g_{S} (a,M_\pi,L) &= c_1 + c_2a + c^{\prime}_3 M_\pi + 
c^{\prime}_4 M_\pi {e^{-M_\pi L}} \,.
\label{eq:CextrapgS}
\end{align}
Adding next order terms such as chiral logs did not improve the fits
(based on the Akaike Information Criteria) and their coefficients were
poorly determined. Variation in estimates on including chiral logs
were, nevertheless, used to obtain first estimates of the possible systematic uncertainty
due to using the lowest order fit ansatz. Our final fits using
Eqs.~\eqref{eq:CextrapgAT}and~\eqref{eq:CextrapgS} are shown in
Fig.~\ref{fig:conUmD_extrap9}. \looseness=-1

The Clover-on-clover esimates on 4 ensembles are consistent with those
from clover-on-HISQ at similar values of the lattice parameters. To
perform analogous fits to obtain results at $a \to 0$ and $M_\pi =
135$~MeV, clover-on-clover calculations are being extended to additional values
of $a$ and $M_\pi$.

\section{Results: Nucleon Charges to quark EDM} 

\begin{figure*}[t]
\begin{center}
\vspace{-0.8cm}
\includegraphics[width=.85\textwidth,trim={0 0 0 2cm},clip]{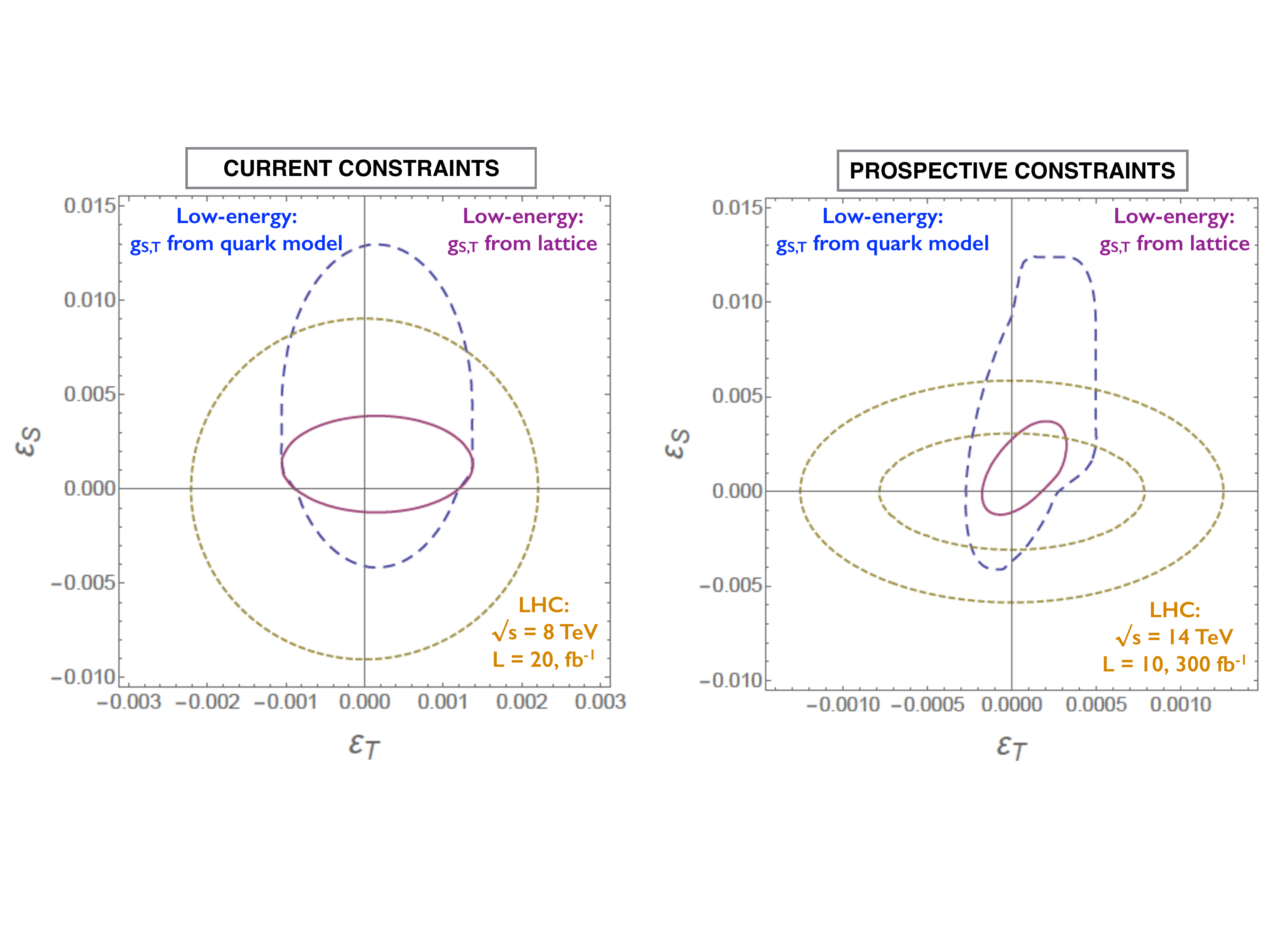}
\end{center}
\vspace{-2.5cm}
\caption{Left panel: current $90 \%$ C.L. constraints on $\epsilon_S$
  and $\epsilon_T$ from beta decays ($\pi \to e \nu \gamma$ and $0^+
  \to 0^+$) and the LHC ($p p \to e \nu + X$) at $\sqrt{s}=8$ TeV.
  Right panel: prospective $90 \%$ C.L. constraints on $\epsilon_S$
  and $\epsilon_T$ from beta decays and the LHC ($p p \to e \nu +
  X$) at $\sqrt{s}=14$ TeV.  The low-energy constraints correspond to $10^{-3}$ 
  measurements of $B,b$ in neutron decay and $b$ in $^6$He
  decay.  Both panels present low-energy constraints under
  two different scenarios for the scalar and tensor charges $g_{S,T}$:
  quark model~\cite{Herczeg:2001vk} (large dashed contour) and lattice
  QCD results given in Eq.~\protect\eqref{eq:results} (small
  solid contour). LHC constraintrs are shown as dotted contours. }
\label{fig:eSeT}
\end{figure*}

\noindent (I) Our results for the isovector nucleon charges, using the
simultaneous fit ansatz defined in Eqs.~\eqref{eq:CextrapgAT}
and~\eqref{eq:CextrapgS} to the 9 clover-on-HISQ data points, are shown in
Fig.~\ref{fig:conUmD_extrap9} and give~\cite{Bhattacharya:2016zcn}
\begin{equation}
g_A^{u-d} = 1.195(33)(20); \qquad g_S^{u-d} = 0.97(12)(6); \qquad g_A^{u-d} = 0.987(51)(20) \,.
\label{eq:results}
\end{equation}
(II) Using the conserved vector current relation $\partial_\mu ({\overline{d}}
\gamma_\mu u)= (m_d-m_u) {\overline{d}} u$, lattice estimates of
$m_d-m_u = 2.67(35)$ given by FLAG~\cite{FLAG:2016qm}, and
our result for $g_S^{u-d}/g_V^{u-d}$ we obtain
\begin{equation}
(M_N-M_P)^{\rm QCD} = g_S^{u-d} (m_d-m_u) / g_V^{u-d} = 2.59(49) \ {\rm MeV}.
\end{equation}
(III) Constraints on novel scalar and tensor couplings, $\epsilon_S$
and $\epsilon_T$, at the TeV scale using low-energy experiments and our $g_S^{u-d}$
and $g_T^{u-d}$ are derived and compared with those
from the LHC in Fig.~\ref{fig:eSeT}.\\ 
(IV) The leading opertors in a low-energy effective
theory that contribute to the neutron electric dipole moment (nEDM)
are the $\Theta$-term, the quark EDM operator and the quark chromo EDM
operators. The ME of the quark EDM operator, same as the flavor
diagonal tensor charges $g_T^{u,d,s}$, are determined to be~\cite{Bhattacharya:2015wna}
\begin{equation}
g_T^{u} = 0.792(42); \qquad g_T^{d} = -0.194(14); \qquad g_T^{s} = 0.007(8) \,.
\end{equation}
In these estimates, the disconnected contributions to $g_T^u$ and
$g_T^d$ have been neglected as they were $O(1\%)$ (smaller than the quoted errors) 
and poorly determined. Using these results and
the experimental bound on the neutron EDM, we performed a first
analysis of constraints on possible quark EDM couplings generated at
the TeV scale and implications for a split SUSY
model in Ref.~\cite{Bhattacharya:2015wna,Bhattacharya:2015esa}. \looseness=-1

\section{Conclusions and Outlook}

Our goal is to calculate the charges and the form factors with
$O(1\%)$ uncertainty on each ensemble and obtain results in the $a\to
0$, $M_\pi =135$~MeV limit with a total error of $2\%$. This will
require simulations with $O(10^6)$ measurements at 4--5 values of the
lattice spacing and on multiple values of the light quark masses close
to the physical pion mass.  To achieve this goal over the next 5--10
years will require further improvements in algorithms for generating
lattices, physics analysis, and the calculation of renormalization
factors. Work towards these three goals is ongoing.

\section*{Acknowledgments}
We thank the MILC Collaboration for providing the 2+1+1-flavor HISQ
ensembles and the JLab/W\&M collaboration for the 2+1 clover
lattices. Simulations were carried out on computer facilities of (i)
Oak Ridge Leadership Computing Facility at the Oak Ridge National
Laboratory, which is supported by the Office of Science of the
U.S. Department of Energy under Contract No. DE-AC05-00OR22725; (ii)
the USQCD Collaboration, which are funded by the Office of Science of
the U.S. Department of Energy; (iii) the National Energy Research
Scientific Computing Center, a DOE Office of Science User Facility
supported by the Office of Science of the U.S. Department of Energy
under Contract No. DE-AC02-05CH11231; and (iv) Institutional Computing
at Los Alamos National Laboratory; and (v) the Extreme Science and
Engineering Discovery Environment (XSEDE), which is supported by the
NSF Grant No. ACI-1053575.  The calculations used the Chroma software
suite~\cite{Edwards:2004sx}. Work supported by the U.S. Department of
Energy, NSF and the LANL LDRD program. \looseness=-1

\vspace{-0.5cm}

\FloatBarrier

\bibliography{ref}
\bibliographystyle{plain}

\end{document}